\documentclass[prl,twocolumn,showpacs,preprintnumbers,amsmath,amssymb,superscriptaddress,floatfix]{revtex4}

\usepackage{graphicx}
\usepackage[T1]{fontenc}
\usepackage{lmodern}
\usepackage[latin1]{inputenc}
\usepackage{dcolumn}
\usepackage{bm}
\usepackage{color}
\usepackage[colorlinks=true,plainpages=false,linkcolor=blue,urlcolor=blue,citecolor=blue,pdfpagemode=UseNone,pdfstartview=FitBH]{hyperref}
\makeatletter\renewcommand{\fnum@figure}[1]{\figurename~\thefigure~(color online).}\makeatother
\definecolor{orange}{rgb}{1,0.7,0}
\definecolor{magenta}{rgb}{1,0,1}

\begin{document}

\title{{Tunable Charge and Spin Order in PrNiO$_3$ Thin Films and Superlattices}}

\author{M.~Hepting$^{1}$, M.~Minola$^{1}$, A.~Frano$^{1,2}$, G.~Cristiani$^{1}$, G.~Logvenov$^{1}$, E.~Schierle$^{2}$, M. Wu$^{1}$, M.~Bluschke$^{1,2}$, E.~Weschke$^{2}$,  H.-U.~Habermeier$^{1}$, E.~Benckiser$^{1}$, M.~Le Tacon$^{1}$ and B.~Keimer}
\email[]{b.keimer@fkf.mpg.de}
\affiliation{Max-Planck-Institut f\"{u}r Festk\"{o}rperforschung, Heisenbergstr. 1, 70569 Stuttgart, Germany\\
$^2$Helmholtz-Zentrum Berlin f\"{u}r Materialien und Energie, Wilhelm-Conrad-R\"{o}ntgen-Campus BESSY II, Albert-Einstein-Str. 15, D-12489 Berlin, Germany}

\date{\today}

\begin{abstract}
We have used polarized Raman scattering to probe lattice vibrations and charge ordering in 12 nm thick, epitaxially strained PrNiO$_3$ films, and in superlattices of PrNiO$_3$ with the band-insulator PrAlO$_3$. A carefully adjusted confocal geometry was used to eliminate the substrate contribution to the Raman spectra. In films and superlattices under tensile strain, which undergo a metal-insulator transition upon cooling, the Raman spectra reveal phonon modes characteristic of charge ordering. These anomalous phonons do not appear in compressively strained films, which remain metallic at all temperatures. For superlattices under compressive strain, the Raman spectra show no evidence of anomalous phonons indicative of charge ordering, while complementary resonant x-ray scattering experiments reveal antiferromagnetic order associated with a modest increase in resistivity upon cooling. This confirms theoretical predictions of a spin density wave phase driven by spatial confinement of the conduction electrons.
\end{abstract}

\pacs{73.21.Cd, 78.30.-j, 75.30.Fv, 75.25.Dk}

\maketitle

The competition between crystalline, liquid, and superfluid states in two-dimensional correlated fermion systems is at the frontier of research on a diverse set of platforms including semiconductor quantum wells, \cite{ShayegPRLan} $^3$He films on surfaces, \cite{Fukuyama} fermionic atoms in pancake optical traps, \cite{Turlapov} and transition metal oxides with quasi-two-dimensional electronic structure \cite{Ghiringhelli}. Artificial superlattices (SLs) of $d$- and $f$-electron compounds have recently opened up the possibility to prepare dense, strongly interacting electron systems in two dimensions \cite{Shishido,Boris2011,Moetakef,Hwang2012,Stemmer}. Tuning the electronic dimensionality through the active layer thickness, and the strength and anisotropy of the interactions through epitaxial strain, affords novel insight into the mechanisms that control the competition between different electronic ground states.

Nickel oxides of composition $R$NiO$_3$ (with $R =$ rare earth other than La) have been of long-standing interest, because the crystallization of the conduction electrons into a charge-ordered state leads to sharp metal-insulator transitions (MIT) \cite{Torrance92,Catalan2008,Alonso2000,Staub02,Scagnoli2005,Acosta-Alejandro2008,Medarde09}. Non-collinear magnetic order of the Ni spins appears at temperatures $T_{\text N} < T_{\text{MIT}}$ \cite{Scagnoli2008,frano13a}. $T_{\text{MIT}}$ decreases and $T_{\text N}$ increases when the single-electron bandwidth is enhanced with increasing size of the $R$ anion, and both transitions coincide for $R =$ Nd, Pr. The compound with the largest anion ($R =$ La) and the largest bandwidth remains metallic and paramagnetic at all temperatures. Recent experiments on nickel-oxide thin films and SLs have demonstrated control of the MIT through dimensionality and epitaxial strain \cite{Boris2011,frano13a,Scherwitzl,Kumah,Liu2013}. Stimulated in part by these results, theoretical research has begun to reassess the traditional interpretation of the charge-ordered state in terms of disproportionation of the Ni$^{3+}$ valence state into Ni$^{3+\delta}$ and Ni$^{3-\delta}$ configurations that are stabilized by Hund's rule interactions \cite{Mazin2007,Blanca2011,Lau,Puggioni,Balachandran2013,Johnston,Jaramillo}. One set of models has invoked an unusual electron-phonon interaction that modulates the covalency of the Ni-O bond \cite{Balachandran2013,Johnston,Jaramillo} so that the charge transfer, $\delta$, between adjacent Ni sites vanishes both above and below the MIT \cite{Johnston}. According to another model, the MIT for $R =$ Nd and Pr is caused by spin density density wave formation, with charge order as a secondary order parameter in some (but not all) lattice structures \cite{Balents1,Balents2}.

Experimental tests of these predictions remain challenging especially for thin films and SLs, where diffraction methods are difficult to apply because of the small scattering signal originating from the atomically thin active layers. Whereas resonant x-ray scattering (RXS) has successfully detected magnetic order in epitaxially strained nickelate SLs and ultrathin films, \cite{frano13a,Liu2013} direct evidence of charge order has not yet been reported in such systems.
%In particular Ni-$K$ edge RXS measurements have successfully shown signatures of charge order in bulk-like $R$NiO$_{3}$ samples \cite{Staub02}, but have given %null results in thin slabs of material.
Prior Raman scattering work on thick films has identified phonon anomalies characteristic of charge ordering, \cite{Zaghrioui,Girardot2008} but in fully strained overlayers of $\sim 10$ nm thickness, the signal from these modes is drowned out by the much more intense Raman spectrum of the substrate. Here we show that this background can be greatly reduced in a carefully aligned confocal Raman spectroscopy setup with polarized, visible laser light. We apply this technique to thin films of PrNiO$_3$ (PNO), the compound with the lowest $T_{\text{MIT}}$ in the $R$NiO$_3$ series, and to SLs of PNO and the wide-band-gap insulator PrAlO$_3$ (PAO). Combined with RXS experiments on the same samples, the results indicate that the theoretically predicted \cite{Balents1,Balents2} spin density wave phase is realized in PNO-PAO SLs.

PNO thin films and SLs were grown by pulsed laser deposition on (001)-oriented substrates of [LaAlO$_{3}$]$_{0.3}$[Sr$_{2}$AlTaO$_6$]$_{0.7}$ (LSAT) with in-plane lattice constant $a_{\textrm{sub}}=3.868$ {\AA} and LaSrAlO$_4$ (LSAO) with $a_{\textrm{sub}}=3.756$ {\AA}. Bulk PNO has a pseudocubic lattice constant of $a_{\textrm{pc}}=3.82$ {\AA}, hence the respective lattice mismatch $(a_{\textrm{sub}}-a_{\textrm{pc}})/a_{\textrm{pc}}$ leads to 1.2 \% tensile strain for PNO on LSAT and to -1.7 \% compressive strain for PNO on LSAO.
%(\textit{approximately the same amount of biaxial strain on compressive and tensile side}).
The films were 12 nm thick, corresponding to thirty-two unit cells (u.c.), and reciprocal-space maps obtained with hard x-rays showed that they were fully strained. The SLs were composed of eight bilayers with four u.c. PNO and four u.c. of PAO. The resistivity was measured via the van-der-Pauw method. The RXS experiments were performed at the BESSY-II undulator beam line UE46-PGM1 at the Ni $L_3$ edge, following a protocol described previously \cite{frano13a}. The intensity of the (1/4, 1/4, 1/4) Bragg reflection (pseudocubic notation) determined by RXS reflects the order parameter of noncollinear antiferromagnetism.

The Raman experiments were conducted with a Jobin-Yvon LabRam HR800 spectrometer using the 632.8 nm excitation line of a HeNe laser. The laser was focused with a $\times 100$ objective, which was positioned with an accuracy $< 0.5$ $\mu$m such that the focus was in the film \cite{SupMat}. Based on a reference measurement with beam focus in the substrate, the substrate contribution was eliminated from the data. Laser heating was minimized by keeping the laser power $< 1$ mW, and the temperatures quoted have an uncertainty of at most 5 K \cite{SupMat}.
%The temperature dependent measurements were carried out in a CryoVac KONTI-micro cryostat attached to the Raman microscope stage.
%More details are given in the Supplemental Material.

\begin{figure}[t]
\includegraphics[width=\columnwidth]{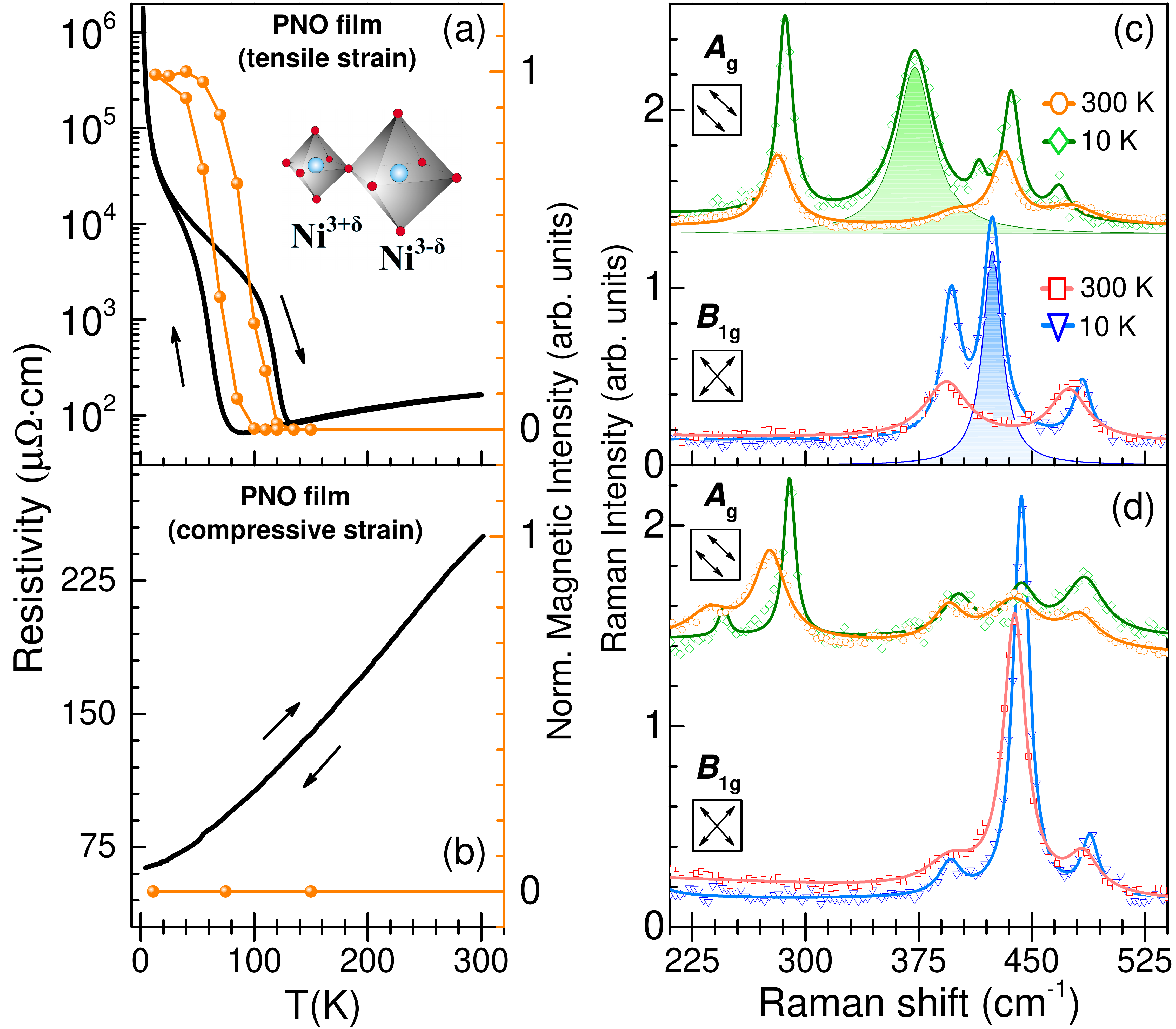}
\caption{\footnotesize{(a,b) Electrical resistivity (left scale) and normalized intensity of the antiferromagnetic (1/4, 1/4, 1/4) reflection (right scale) \cite{frano13a} measured by RXS as a function of temperature in PNO films under (a) tensile and (b) compressive strain. Arrows indicate warming and cooling cycles. (c,d) Raman spectra of PNO films under (c) tensile and (d) compressive strain at $T = 300$ and 10 K in A$_{\textrm{g}}$ and B$_{\textrm{1g}}$ symmetry (upper and lower subpanels, respectively). The former film exhibits anomalous modes at low $T$ associated with charge ordering (shaded in panel c), while the latter does not. The solid lines represent the results of least-squares fits of the data to Voigt profiles. The inset in (a) shows a segment of the charge ordering pattern with two differently sized NiO$_6$ octahedra. The arrows in the insets in (c,d) indicate the electric field vectors of the incoming and the scattered light with respect to the pseudocubic unit cell of PNO.}}
\label{fig:figure1}
\end{figure}

Figure \ref{fig:figure1} shows the Raman spectra of 12 nm thick PNO films on tensile and compressive strain obtained in this way, together with the electrical resistivity and the intensity of the RXS Bragg reflection characteristic of the antiferromagnetic order parameter measured on the same samples as a function of temperature. The film under tensile strain shows a MIT with significant hysteresis in the temperature range 80-130 K, which is indicative of a first-order transition (Fig. \ref{fig:figure1}a). The onset of magnetic order determined by RXS coincides approximately with the MIT, as it does in bulk PNO \cite{Torrance92}. By contrast, the compressively strained PNO film remains metallic and paramagnetic down to lowest temperatures (Fig. \ref{fig:figure1}b). The dependence of the MIT on epitaxial strain in PNO is qualitatively similar to recent findings in NdNiO$_3$ (NNO) thin films \cite{Liu2013}.

\begin{figure}[t]
\includegraphics[width=\columnwidth]{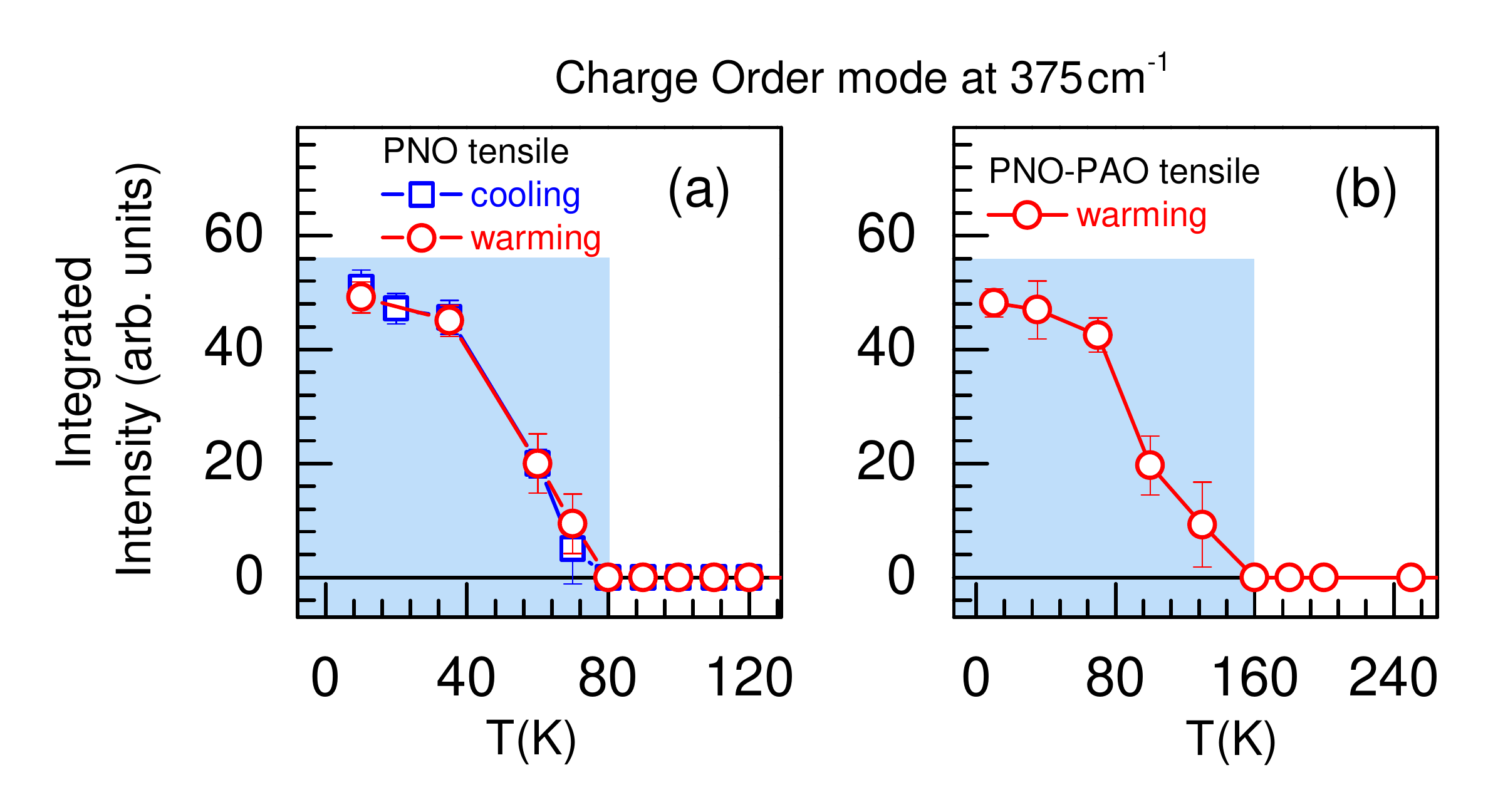}
\caption{\footnotesize{Temperature dependence of the intensity of the anomalous phonon mode at 375 cm$^{-1}$ in (a) the PNO film and (b) the PNO-PAO SL under tensile strain. No hysteresis is observed. In the shaded areas, the film and the SL are insulating.}}
\label{fig:figure2}
\end{figure}

\begin{figure}[t]
\includegraphics[width=0.7\columnwidth]{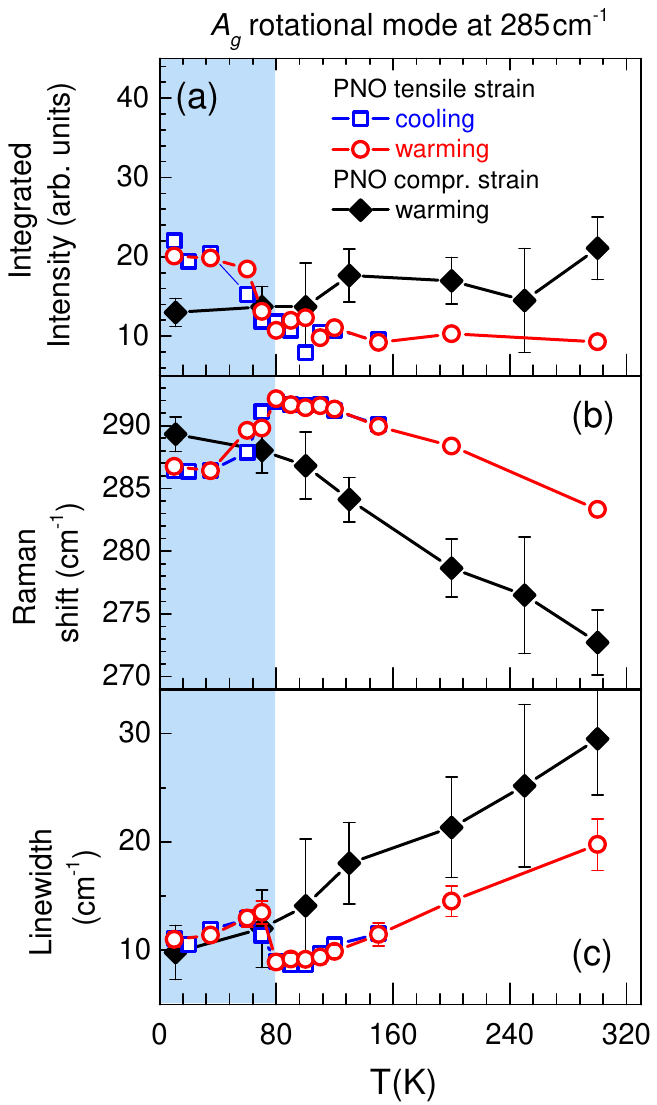}
\caption{\footnotesize{Temperature dependence of the (a) intensity, (b) frequency, and (c) linewidth of the regular phonon mode at 285 cm$^{-1}$ of PNO films under tensile (open symbols) and compressive (closed symbols) strain. In the shaded areas, the film under tensile strain is insulating.}}
\label{fig:figure3}
\end{figure}

The Raman spectra displayed in Figs. \ref{fig:figure1}c,d supply complementary information about charge ordering, which modifies the lattice structure and hence the phonon spectrum. At high temperatures, the Raman-active phonons reflect the pseudocubic perovskite structure, with uniform Ni valence. Upon cooling below $T_{\text{MIT}}$, the spectrum of PNO under tensile strain exhibits two additional modes at 375 cm$^{-1}$ and 425 cm$^{-1}$ (Fig. \ref{fig:figure2}c), consistent with a superstructure of two differently sized NiO$_6$ octahedra, which in the bulk leads to a loss of the $b$ mirror symmetry of $Pbnm$ and to the monoclinic space group $P2_1/n$ \cite{Alonso2000,Piamonteze2003,Zaghrioui,Girardot2008}. The association with charge ordering is supported by the absence, at low temperature, of additional phonons in the compressively strained PNO film, which does not exhibit a MIT (Fig. \ref{fig:figure2}d). Polarization-resolved data in different light-field configurations (indicated in Figs. \ref{fig:figure1}c,d as arrows relative to the pseudocubic unit cell of PNO) determine the symmetry of the displacement patterns as $A_{\textrm{g}}$ and $B_{\textrm{1g}}$ for the modes at 375 cm$^{-1}$ and 425 cm$^{-1}$, respectively.

The Raman spectra were fitted to a superposition of Voigt profiles resulting from a convolution of the intrinsic Lorentzian lineshape with the Gaussian spectrometer resolution (lines in Figs. \ref{fig:figure1}c,d). Figure \ref{fig:figure2}a shows the integrated intensity of the anomalous mode at 375 cm$^{-1}$ characteristic of charge ordering obtained in this way. Its temperature dependence exhibits behavior characteristic of the order parameter of a continuous (or weakly first-order) phase transition with critical temperature $\sim 80$ K. Figure \ref{fig:figure3} displays the  intensity, frequency, and linewidth of a regular mode with frequency 285 cm$^{-1}$ that corresponds to rotations of the NiO$_6$ octahedra in the pseudocubic perovskite structure. All of these parameters exhibit anomalies at the same temperature, reflecting changes in the electron-phonon coupling at the MIT; similar observations were reported before for $R =$ Nd and Sm \cite{Zaghrioui,Girardot2008}. In contrast, the regular phonons in the compressively strained PNO film exhibit standard anharmonic behavior, without anomalies.

In the PNO film under tensile strain, we carefully monitored the phonon lineshapes as a function of temperature while eliminating laser heating effects \cite{SupMat}. Remarkably, we did not find any hysteresis in any of the phonon lineshapes upon cycling through the MIT (Figs. \ref{fig:figure2}a and \ref{fig:figure3}), in stark contrast to the behavior of the resistivity and the magnetic order parameter measured on the same sample (Fig. \ref{fig:figure1}a). Differences in the hysteretic behavior of the regular Raman phonons and the macroscopic resistivity have been noted before, and were attributed to the possible influence of defects on $dc$ charge transport \cite{Girardot2008}. Similar considerations were made in interpreting differences between the macroscopic magnetic susceptibility and the $dc$ resistivity \cite{Kumar2013}. The results of our experiments now allow us to compare data generated by two microscopic probes of magnetic and charge order, respectively. The fact that the former persists to $\sim 120$ K upon heating (Fig. \ref{fig:figure1}a), while the latter vanishes at $\sim 80$ K (Figs. \ref{fig:figure2}a and \ref{fig:figure3}), suggests that the two order parameters are only weakly coupled, and that the sublattice magnetization is the primary order parameter. This is consistent with the theoretically predicted spin density wave phase \cite{Balents1,Balents2}.

\begin{figure}[t]
\includegraphics[width=\columnwidth]{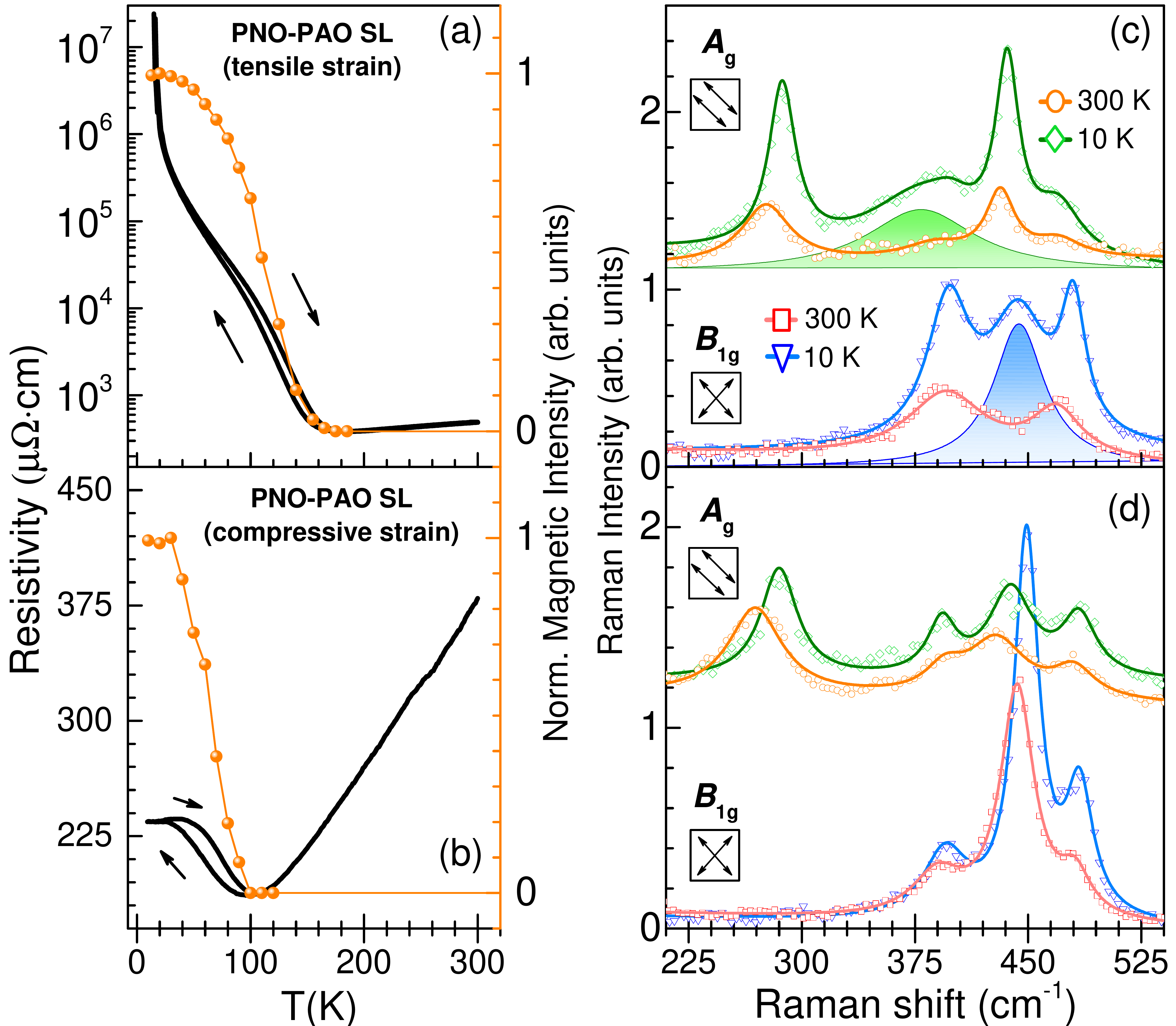}
\caption{\footnotesize{(a,b) Electrical resistivity and normalized intensity of the antiferromagnetic (1/4, 1/4, 1/4) reflection measured by RXS as a function of temperature in PNO-PAO SLs under (a) tensile and (b) compressive strain. The arrows indicate warming and cooling cycles. (c,d) Raman spectra of PNO films under (c) tensile and (d) compressive strain at $T$ = 300 and 10 K for the A$_{\textrm{g}}$ and B$_{\textrm{1g}}$ channels. The anomalous modes at low \textit{T} (shaded in (c)) are due to charge order. Lines and symbols in the insets are analogous to Fig. 1.}}
\label{fig:figure4}
\end{figure}

We now turn to the measurements on PNO-PAO superlattices grown under the same strain conditions as the PNO films. Figure \ref{fig:figure4} shows that the spatial confinement of the conduction electrons in the SLs strengthens the magnetic correlations. In the SL under tensile strain, $T_{\text N}$ grows to $\sim 160$ K, (Fig. \ref{fig:figure4}a) while the hysteresis that characterizes the transition is greatly reduced compared to the corresponding film (Fig. \ref{fig:figure1}a). In the SL under compressive strain, the spatial confinement induces antiferromagnetic order with $T_{\text N} \sim 100$ K, (Fig. \ref{fig:figure4}b) in a manner similar to SLs based on LaNiO$_3$, \cite{Boris2011} but already at an active layer thickness of four (rather than two) unit cells. The resistivity upturn upon cooling below $T_{\text N}$ of the compressively strained PNO-PAO SL is much more subtle than in the PNO films and in the SL under tensile strain, and the SL remains highly conductive even below $T_{\text N}$. Combined with the RXS data that demonstrate similar magnetic order parameters in both samples, this suggests that differences in charge order may be responsible for the different transport properties.

The Raman spectra of the SLs provide direct insight into the influence of spatial confinement and strain on charge ordering. Figure \ref{fig:figure4}c,d presents polarized spectra of the PNO-PAO SLs, in a manner parallel to those of the films (Fig. \ref{fig:figure1}c,d). The phonon profiles in the SLs are generally broader than in the films, presumably due to unresolved sidebands resulting from hybridization of vibrations in the PNO and PAO subunits and/or disorder. Apart from this broadening, however, the energies and polarization dependences of the regular phonons in both SLs are closely similar to those of the corresponding films. In particular, the modes at 285 cm$^{-1}$, 435 cm$^{-1}$ and 465 cm$^{-1}$ in the SL under tensile strain only appear in the $A_{\textrm{g}}$ channel while the ones at 400 cm$^{-1}$ and 480 cm$^{-1}$ only appear in $B_{\textrm{1g}}$. The regular modes in the spectra of both SLs exhibit the standard anharmonic hardening and narrowing upon cooling.

Whereas only such regular modes appear in the spectrum of the SL under compressive strain, anomalous modes closely similar to those in the film under tensile strain (Fig. \ref{fig:figure1}c) also appear in the SL under tensile strain at 375 cm$^{-1}$ in $A_{\textrm{g}}$ and at 445 cm$^{-1}$ in $B_{\textrm{1g}}$ configurations, respectively (shaded areas in \ref{fig:figure4}c). The energy of the $B_{\textrm{1g}}$ mode is $\sim$ 20 cm$^{-1}$ higher than the one of the corresponding PNO film, and both $A_{\textrm{g}}$ and $B_{\textrm{1g}}$ phonons are significantly broader in the SLs (mirroring the broadening of the regular modes), but their overall integrated intensities are similar to those of the film. The onset temperature of the anomalous modes (Fig. \ref{fig:figure2}b) coincides with $T_{\text{MIT}}$ within the experimental error, once again strengthening the association with charge order. Their absence in the spectra of the SL under compressive strain (Fig. \ref{fig:figure4}d) therefore implies that the amplitude of charge modulation is either absent or much smaller than in the SL under tensile strain.

In summary, we have used two complementary, microscopic probes to elucidate the influence of temperature, epitaxial strain, and spatial confinement on charge and spin order in PrNiO$_3$. The RXS data show that both SLs have identical magnetic ground states, while our Raman scattering data imply that the amplitude of the charge order differs greatly in both systems. This is consistent with the prediction by Balents and coworkers \cite{Balents1,Balents2} that the spin density modulation is the primary order parameter, whereas a charge density modulation may or may not accompany the spin density as a secondary order parameter, depending on the specific lattice symmetry.

We have thus identified three different states, only two of which were observed before in bulk $R$NiO$_3$: an insulating state with robust spin and charge order in films and superlattices under tensile strain, and a metallic state with neither form of order in films under compressive strain. In superlattices under compressive strain, a weakly metallic state with fully developed spin order but no (or very weak) charge order is realized. Further evidence of decoupling of spin and charge order comes from films under tensile strain, where we showed that superheating of magnetic order persists over a much larger temperature range than charge ordering. In contrast, we find no evidence for a charge-ordered ``spin liquid'' state with converse combination of both order parameters \cite{Liu2013}.

Our data thus directly confirm theoretical predictions for a spin density wave phase with charge order as a secondary order parameter \cite{Balents1,Balents2}. The mechanism invoked by these predictions, Fermi surface nesting, is enhanced by spatial confinement, in accord with our data on PNO films and PNO-PAO SLs under the same (compressive) strain conditions. Spin density waves have been directly observed and characterized only in few bulk metals, including elemental chromium and its alloys \cite{Fawcett}, organic charge-transfer salts \cite{Gruner}, and most recently the iron pnictides \cite{Stewart}. The separate control of magnetic and charge order we have reported here for the nickelates opens up new opportunities for device applications, including designs in which metallic antiferromagnets serve as active elements \cite{MacDonald,Sinova}. In addition, our determination of the energies and symmetries of the vibrational modes characteristic of charge ordering, as well as their dependence on strain and spatial confinement, provides specific input for models of charge ordering and electron-phonon interactions in the nickelates and related transition metal oxides \cite{Mazin2007,Blanca2011,Lau,Puggioni,Balachandran2013,Johnston,Jaramillo}.

Finally, we have demonstrated that confocal Raman spectroscopy with visible laser light \cite{SupMat} is a powerful alternative to ultraviolet Raman spectroscopy \cite{Tenne1} for the investigation of lattice vibrations in epitaxially strained thin films and superlattices. We have also shown that the technique is accurate enough to resolve phonon anomalies associated with electronic phase transitions in thin films and heterostructures. We thus foresee manifold applications in the determination of the structural and electronic phase behavior of oxide heterostructures.

M. H. and M. M. contributed equally to this work. We thank A. V. Boris, Y. Lu, C. Dietl, G. Sawatzky, and J. M. Triscone for fruitful discussions. We acknowledge financial support from the German Science Foundation under Grant No. TRR80.

%\end{thebibliography}

\end{document}